\documentstyle[wstwocl]{article}
\begin{document}

\bibliographystyle{unsrt}    

\newcommand{\st}{\scriptstyle}
\newcommand{\sst}{\scriptscriptstyle}
\newcommand{\mco}{\multicolumn}
\newcommand{\epp}{\epsilon^{\prime}}
\newcommand{\vep}{\varepsilon}
\newcommand{\ra}{\rightarrow}
\def\la{\lambda}
\def\be{\begin{equation}}
\def\ee{\end{equation}}
\def\bea{\begin{eqnarray}}
\def\eea{\end{eqnarray}}
\def\CPbar{\hbox{{\rm CP}\hskip-1.80em{/}}}
\def\yuk{Yukawa coupling}
\def\sus{supersymmetr}
\def\sug{supergravity}
\def\SM{Standard Model}
\def\bi{\bar{\imath }}
\def\bj{\bar{\jmath }}

\setcounter{secnumdepth}{2} 

   
\title{FCNC IN SUSY THEORIES}

\firstauthors{C. A. Savoy}

\firstaddress{Service de Physique Th\'eorique, C.E. de Saclay, 
91191 Gif-s/Yvette, France}

\secondauthors{}


\secondaddress{}


\twocolumn[\maketitle\abstracts{ 
Recent work on flavour changing neutral current effects in \sus ic
models is reviewed.  The emphasis is put on new issues related to solutions
to the flavour problem through new symmetries: GUTs, horizontal symmetries,
modular invariances. \vskip 9pt
\hskip 65pt \underline{{\sl Invited Talk at the
HEP95 Euroconference, Brussels, July 95.}}}]

\section{Introduction}
A  rich literature  is available 
about FCNC restrictions on \sus ic extensions of the standard 
model. Nevertheless, both the LEP (and Tevatron) constraints on \sus ic 
theories and some fresh insight on spontaneously broken supergravities from 
superstrings have encouraged a recent revival of this subject.

The basic \sus y induced FCNC (SFCNC) effects are produced by the 
analogues of the \SM \   loop diagrams for neutral 
current processes, with quarks and vector bosons 
replaced by squarks and gauginos.  If quark(lepton) 
and squark(slepton) mass matrices are not 
diagonal in the same basis, even the the couplings to neutral gauginos to fermions and sfermions will not be diagonal and will induce 
FCNC effects.  There are several sources of flavour mixing in 
gaugino couplings that we now turn to discuss. However, 
I want to keep in mind that \sus y must be a local symmetry, 
namely, a \sug \  theory, at the fondamental level. This has 
implications on the structure of the low energy effective theory
(and vice-versa, which is even more important!)

Within the general framework of \sug ,  
a theory is defined by the gauge and matter superfields, 
and by their couplings encoded in 
the K\"ahler potential and the superpotential. The
low-energy theory is then fixed by the values of the auxiliary fields that
provide \sus y breaking and their couplings
to the light fields. The \sus ic part of this effective theory contains 
the \sus ized gauge couplings and the \sus ized Yukawa couplings,
encoded in an effective superpotential
$W=$$\sum[ Y_{ij}^UQ^iU^jH_2+Y_{ij}^DQ^iD^iH_1+
Y_{ij}^E L^iE^jH_1] ,$ where $H_1$ and $H_2$ are the two Higgs superfields,
and the matter superfields are as follows: $Q, 
L,$ contain the ${\rm SU(2)_{weak}}$ doublets of quarks and
leptons, and $U,D,E$ contain the right-handed quarks and leptons. The
physical content of the three Yukawa coupling $3\times 3$ matrices      
is given by their eigenvalues $Y_i(i=u,c,t;d,s,b;e,\mu ,\tau )$ as
well as the CKM matrix $V$. The observed
quark masses and mixings and lepton masses reveal a strong hierarchy  
conveniently displayed in terms of a small parameter which we choose
to be the Cabibbo angle, $\lambda =.22$: $Y_t:Y_c:Y_u=\la ^8:\la ^4:1,
Y_b:Y_s:Y_d=\la ^4:\la ^2:1, Y_{\tau}:Y_{\mu}:Y_e=\la ^4:\la ^2:1,
V_{us}=\la , V_{cb} \sim  \la ^2, V_{ub} \sim  \la ^3 $.

At the level of the effective theory, below the Planck scale, 
the \sus y breaking effects reduce to gaugino masses and the soft 
interactions in the scalar potential.  The 
scalar $ {\rm (mass)}^ {\rm 2} $ matrix depend on the K\"ahler 
potential and on the \sus y breaking auxiliary fields.  
The universality or flavour 
independence hypothesis assumes equal masses for all squarks at the 
unification. At lower energies, radiative corrections from Yukawa interactions split this degeneracy  with flavour dependent shifts. The triscalar 
couplings are basically proportional 
to couplings in the superpotential. Again, if  universality is 
assumed for the proportionality factors, referred to as $ A $-parameters,
their equality is spoilt at lower energies by the 
calculable radiative corrections.

Universality 
of soft terms is often assumed in SFCNC studies. Then, the most 
striking effects of 
 radiative corrections are of two kinds. Gauge corrections are almost
universal and attenuate loop effects by an overall rise 
in the squark masses if gluinos are relatively heavy. Yukawa corrections dominated by the top coupling, $ Y_ t, $ tend to align the down squark 
mass eigenstates to the up quarks (if $\tan{\beta}$ is not too large).
 This reverses the pattern of 
gaugino couplings in comparison with the gauge boson ones. Chargino couplings 
to down squarks and up quarks are approximately flavour diagonal while gluino 
and neutralino couplings become proportional to the CKM matrix. 
However, the expected physical effects are either consistent with the 
present overall bounds on \sus ic particles or they depend on unknown
mixings and phases, but the $ b \rightarrow s\gamma $
transition provides interesting information.

Thus, universality naturally suppresses SFCNC effects as it amounts to 
postulate the largest possible horizontal
symmetry,  ${\rm U(3)^5}$, for each of the 5 irreps of the \SM \ 
in the 3 fermion families, as an accidental symmetry, {\sl i.e.}, a symmetry 
of the scalar potential in the limit where all \yuk s vanish. 
This is justified if \sug y couplings
to the \sus y breaking are flavour independent. As we now turn to discuss,
they are not necessarily so.

\section{Flavour theories and \sus y}
The fermion unit in the \SM \  is a  family of 15 fermions that
provide a non-trivial anomalous-free representation of the gauge 
group. GUTs are attempts to understand the fermion pattern by 
(vertical )unification of the elements of the family within
a representation of a larger gauge theory at very high energies. 
The triplication of families is a puzzle. But these fermion 
replicas do not look as clones since they quite differ by the 
strong hierarchy in  their \yuk s.  The natural
explanation of this situation is to hypothetize that quarks/leptons of the
same charge have different quantum numbers of some new symmetries at high energies \ (symmetries that commute with the \SM symmetries have been
called horizontal). 

As in many particle physics issues, hints come from superstrings models,
where one finds examples of compactifications with  fermion
families  and neither vertical nor horizontal
unification. Instead, there are in general additional abelian $ U(1) $ 
symmetries that differentiate between fermions. 
Moreover, the superstring theory particle masses and
couplings are field dependent dynamically determined quantities.

A conspicuous result of superstring studies is that the three families 
of quark superfields may couple to \sug \ according to different 
terms in  the K\"ahler potential. The relevant low energy limit of 
superstring models are described by
a $N=1$ suoergravities. The zero-mass string spectrum contain 
an universal dilaton 
$S $, moduli fields, related to the compactification of six superfluous 
dimensions, denoted by  $T_{\alpha} (\alpha =1..m),$ 
 and matter chiral fields $A ^i.$ A crucial role is 
 played by the target-space modular
symmetries $SL(2,Z)$ , transformations on the  $T_{\alpha}$  that are
invariances of the  effective supergravity theory.  In string models
 of the orbifold type, the matter fields $A^i $ transform under $SL(2,Z)$ 
according to a set of numbers, $n_i^{(\alpha )}$,
 called the modular weights of the
fields $A^i$ with respect to the modulus $T^\alpha .$  

 The dilaton superfield in these theories does have universal 
\sug \ couplings to matter superfields. But the moduli  
couplings are fixed by modular invariances. Thus, the K\"ahler
potential and the superpotential can have different dependences on the
moduli for each flavour.  On the other hand, these moduli correspond
to flat directions of the scalar potential so that their vev's are
fixed by quantum corrections. Assuming that the relevant ones come
from the light sector, namely by the coupling of moduli to quarks
and leptons in the low energy theory, it  has been suggested that
modular invariances can also provide a theory of flavour, by predicting 
the hierarchies in the moduli dependent \yuk s. This interesting 
idea is discussed in more detail in the contributions\cite{Du} of E. Dudas
and F. Zwirner.
For this reason, it is not developped here.  

Motivated by superstrings, as well as symmetries proposed to explain the 
structure of Yukawa couplings, new analyses\cite{BIM,NS} have been performed
on FCNC transitions produced by non-universality in \sug
couplings. Of course, the results are model dependent, one variable
being the amount of the flavour independent \sus y breaking 
(in the dilaton direction) responsible for gaugino masses,
that attenuates SFCNC. With this proviso the more important constraints
in the quark sector are coming from K-physics. The lepton sector is less sensitive to gaugino masses, and lepton flavour violations put
severe constraints on the parameters, but only as functions of
 unknown lepton mixing angles.
   
Nevertheless, in this talk I would like to focus on the SFCNC problem 
from the stand-point of different attempts to explain the origin 
of flavour, hence of fermion masses and mixings.   

\section{SFCNC effects from SUSY unification}
Recently, the question of FCNC effects arising from SUSY GUTs has
been analysed in detail in a series of papers\cite{BHS} . This
possibility was pointed out already some time ago, but the fact 
that the top \yuk \  is so large considerably enhances the 
effects. The idea is to estimate the renormalization correction
from the running of the soft parameters in the theory from the
\sug scale ($M_{Planck}$) down to the GUT scale ($M_{GUT}$)
in presence of very large \yuk s, which is certainly the case for
$Y_t.$ In a GUT, above $M_{GUT}$, the following part of the superpotential 
give also rise to loop diagrams 
$\sum[ Y_{ij}^UE^iU^jH_3$ $+Y_{ij}^DQ^iL^jH'_3$ $+
Y_{ij}^D D^iE^jH'_3]$ involving the Higgs triplet partners. 
The coupling $Y_t$ is allways large,
while $Y_b=Y_{\tau}$ is large in O(10) unification or even for SU(5)
with large $\tan{\beta}.$ The effect of the running from $M_{Planck}$
to $M_{GUT}$ can be very important: the $\widetilde{\tau} _R $
 is roughly reduced by a factor
$(1-Y^2_t/2Y^2_{max}),$ defined at $M_{GUT},$ where $Y^2_{max}$ is the
value of $Y_t$ for a Landau pole at $M_{Planck}.$ The mass splitting with
 respect to $\widetilde{e} _R $ and $\widetilde{\mu} _R $ will remain 
 at low energies and produce lepton flavour violating processes. Of course the
results also depend on the angles defined by the diagonalizattion of
the lepton and slepton masses. Assuming naive GUT relations for the
lepton mixings - {\sl cum grano salis} in view of the bad naive GUT 
predictions for the two light families - one gets sizeable FCNC 
effects in large regions of the parameter space. For large
$Y_b$ the effects are even bigger. The results can be illustrated 
by assuming universal boundary conditions
at $M_{Planck}$, so that the slepton splitting is only due to the 
Higgs triplet. In this case, it is possible to present
detailed predictions for the various lepton flavour violating processes
(for quark FCNC, those are concealed by the analogous
contributions from the MSSM superpotential).

Of course if one attempts a real theory of fermion masses based on 
GUTs, and O(10) has been preferred in this respect\cite{DHR}, 
for instance, by the introduction of non-renormalizable interactions 
and discrete symmetries,
there will be corresponding constraints on the soft
scalar masses and couplings. The framework will be similar to what
is discussed herebelow in the case of abelian horizontal symmetries.

\section{The pseudo-Goldstone approach}
Dimopoulos and Giudice\cite{DGT} invoke the pseudo-Goldstone
phenomenon to enforce FCNC suppression. They assume a large 
$\Pi_{A=Q,U,D,L,E)}U(3)$'accidental' symmetry of the scalar potential, 
including the scalar masses, in the limit of vanishing \yuk s. 
They introduce on-purpose multiplets, say in the $ Adj(U(3)^5),$ 
whose vev's break 
$ U(3)^5 \rightarrow U(1)^{15} $ or $\rightarrow [U(2) \times U(1)]^5. $
The remaining symmetries entail the following form for each one of the
sfermion mass matrices:\ $\widetilde{m}^2 _A=$ $e^{-i\theta_A}diag( 
\widetilde{m}^2 _{A1}, \  \widetilde{m}^2 _{A2} \   
\widetilde{m}^2 _{A3} )e^{i\theta_A},$ where $\theta_A$ are matrices, each one containing
five Goldstone fields living in the coset $ U(3)/U(1)^3 $ ( the extension
to $[U(2) \times U(1)]$ is obvious). These are massless states as the
potential is flat along the $\theta_A$ directions.    
Actually, they are 'pseudo-Goldstone' states since the flavour
symmetries are explicitly broken by the \yuk s. The latter are taken
{\sl a priori} as given by the quark masses and CKM mixings. Then, at the
quantum level, the hidden flavour symmetry is broken by loops with quarks
that spoil the flatness along the $\theta_A$ directions. By minimization
one obtains the $\theta_A$ vev's (and masses) in terms of the \yuk s 
$Y _A ,$ such that the  $\widetilde{m}^2 _A$'s are all aligned 
to the \yuk s $Y_A$ but $\widetilde{m}^2 _Q$ to the matrix $Y ^2_U+
K^{+}Y ^2_D K $. The quark squark alignment is as good as possible,
still the $\widetilde{m}^2 _Q$ disalignment could induce 
too much $ K\bar{K} $ mixing.
This is avoided if the remaining accidental symmetry is 
$U(2) \times U(1) $ so that $\widetilde{m}^2 _{Q1}=\widetilde{m}^2 _{Q2}.$
This can be implemented\cite{DGT} by
enlarging the accidental $ U(3)^5 $ symmetry to $O(8)$, 
spontaneously broken into $O(7).$
  
In spite of its formal elegance, this approach does not address the 
flavour problem as far as the expected dependence of the \yuk s on new
fields is not envisaged while it might provide a prediction for quark
masses as well. Also, the necessarily large number of {\sl ad hoc}
Goldstone fields could mitigate one's enthusiasm.

\section{The \sus ic Froggatt-Nielsen approach}
The smalness of the mass ratios and mixing angles faces us 
with a problem of naturalness. The direction initiated by Froggatt and 
Nielsen~\cite{FN} to understand such a hierarchical pattern goes as follows:  
 ({\sl i}) The key assumption is a gauged horizontal $U(1)_X$ 
 symmetry violated by the small quark masses so that 
small Yukawa couplings are protected by this symmetry. 
The effective $U(1)_X$ symmetric theory below
some scale $M$ is supposed to be natural to the extent that all parameters
are of $O(1)$. The scale $M$ is the limit of validity of the effective
theory, of $O(M_{Planck})$ if one adopts a superstring
point of view. The X-charges of quarks, leptons and Higgses are 
free parameters to be fixed
{\sl a posteriori} and simply denoted $q_i$,$u_i$,
$d_i$,$l_i$,$e_i$,$h_1$,$h_2$, for the different flavours, where $i=1,
2,3$  is the family index.   
({\sl ii})  One (or more) Froggatt-Nielsen 
field $\Phi ,$ a \SM gauge singletis introduced,  and we normalize 
the $U(1)_X$ so that its charge is $X_{\Phi} = -1$. The effective (non-renormalizable) 
$U(1)_X$ allowed couplings are then of the form 
$g^U_{ij}(\Phi /M)^{q_i + u_j + h_2} Q^i U^j H_2 $, with analogous 
expressions for the  $H_1$ couplings to down quarks and leptons. The 
coefficients $g^U_{ij}$, etc, are taken to be  natural, {\sl i.e.}, 
of $O(1)$, unless they are required to vanish by the $U(1)_X$ symmetry.
 ({\sl iii})  The small parameter $\la $ is identified with the
ratio $({<\Phi> \over M})$ as the $U(1)_X$ symmetry is broken by
the $\Phi $ v.e.v.. Below the scale $<\Phi> = \la M$, one recovers
the \SM \ with the effective Yukawa coupling matrices given by
$ Y^U_{ij} =  \la^{|q_i+u_j+h_2|} ,$ $ Y^D_{ij}=\la^{|q_i+d_j+h_1|}$,
$Y^E_{ij}= \la^{|l_i+e_j+h_1|}.$ The Yukawa matrix entries
corresponding to negative total charge should vanish but these zeroes are 
filled by the diagonalization of the $\la$ -dependent metrics.   

The X-charges are now chosen to fit the hierarchy in the mass eigenvalues
and mixing angles. The experimental masses
(at $O(M_{Planck})$) of the third families give: 
$h_2+q_3+u_3=0$ and 
$x=h_1+q_3+d_3=h_1+l_3+e_3,$     
where the parameter $x$ depends on the assumed value for $\tan \beta.$ 
With this restriction the Yukawa couplings
depend only on the  charge differences $q_i - q_3,$
 $u_i - u_3,$..., $e_i-e_3 $ and $x$. 

Recently,
there has been an intensive investigation of this model\cite{IR,DPS1}, 
including a classification of the possible charge assignments\cite{DPS1}.
But the question I would like to discuss here was first investigated by 
Leurer, Nir and Seiberg\cite{NS} in the Froggatt-Nielsen framework.
Just like the \yuk s, the soft \sus y breaking terms contain powers of 
the $\Phi$ -field to implement the  $U(1)_X$ symmetry. The scalar
mass matrices have a corresponding hierarchy among their 
elements, so that 
$(\widetilde{m}^2_A)_{i{\bj }}= $ $f_{Aij}\la ^{|q_i - q_j|},$ A=Q,U,D,L,E, 
where, in the absence of any other symmetry
principle, the coefficients $f_{Aij}$ are all of the order of the 
\sus y breaking parameter $m_{3/2}^2$, where $m_{3/2}$ is the gravitino mass. 
Even in the flavour basis that diagonalizes quark mass matrices,
the squark mass matrices will still be of the same non-diagonal form.
Therefore large FCNC effects might be induced from loop diagrams with
the exchange of neutral sfermions (gluino, photino,...) in possible 
 disagreement with experiments. Indeed, with only one 
$\Phi$ -field , the acceptable $U(1)_X$ 
charge asignements yield  $(\widetilde{m}^2_D)_{12}(\widetilde{m}^2_Q)_{12}$
$\propto {{m_d} \over {m_s}}$, which imply much too large FCNC effects in
K-physics. One solution\cite{NS} is to double the Froggatt-Nielsen, with 
another abelian 
symmetry and a smaller scale. In this case it is possible to strongly suppress
$(\widetilde{m}^2_D)_{12}$. Interestingly enough, the model predicts large 
$(\widetilde{m}^2_U)_{12}$ leading to sizeable $D\bar{D}$ mixing that could 
be experimentally tested.

Another solution\cite{DPS1} is to assume only one more singlet 
$\Phi '$  and an appropriate charge asignment so that $(\widetilde{m}^2_D)_{12}(\widetilde{m}^2_Q)_{12}$
$\propto {{m_d^2} \over {m_s^2}}$, which is just enough. Remarkably,
in this model all anomalies related
to $U(1)_X$ can be cancelled, while in the other models one has to rely
upon the Green-Schwarz mechanism\cite{IR,DPS1}.

\section{Horizontal symmetries in \sug }
On one hand, horizontal symmetries are a natural way to solve
the family puzzle and the fermion mass hierarchy, and give some 
restrictions on squark masses as well. On the other hand,
 in string inspired \sug , the sfermion masses depend on the 
modular properties of the matter fields and their modular 
dependence might well be related to the origin of flavour.
What if one imposes both symmetries on a broken \sug \  model?
This has been recently investigated\cite{DPS2}. For definiteness,
let us define the modular properties by
some set of modular weights $n_i^{(\alpha )}$ associated to 
each of the matter fields, and their transformation under an
abelian $U(1)_X$ symmetry implementing the Froggatt-Nielsen
 mechanism, by their  charges $X_i$. Analogously, $ n_{\Phi} ^{(\alpha )}$
and $X_{\Phi} $ are introduced for the singlet field $\Phi.$ 
Now, let us require 
the \sug \  theory to be invariant under these $SL(2,Z)$
and $U(1)_X$ transformations. Then, one shows the very
interesting relation: $(q_i-q_j)n_{\Phi}^{(\alpha )} =$ $X_{\Phi} 
(n_{q_i}^{(\alpha )}-n_{q_j}^{(\alpha)})$ between charge and modular 
weight differences. Though the results are easily generalized\cite{DPS2},
let us keep only one modulus, say, the overall one, $T$.
Through some mechanism that we do not quite
understand yet, the dilaton $S$ and the moduli $T$
get their vev's that fix the gauge couplings and the compactified dimensions in string theory. Then, assume \sus y is broken by the auxiliary 
components of the $S$ and $T$ supermultiplets, $F_S$ and $F_T$, and define the 
so-called gravitino angle\cite{BIM}, $\tan{\theta }={F_S}/{F_T} .$
The $\Phi$ vev, in this one-singlet case, is fixed by the Fayet-Iliopoulos
term to be of $O(\la M_{Planck})$, and the \sus y breaking is precisely
fixed in terms of $ n_{\Phi}$  and $X_{\Phi}$, with a $F_{\Phi}$ and a $D_X$
components. Then the squark and slepton  masses can be calculated, with
a surprisingly simple expression, resulting of the coalescence of
all sources of \sus y breaking. For instance, for diagonal entries one
gets the relations: $\widetilde{m}_{i\bi }^2-\widetilde{m}_{j\bj }^2 =$ 
$(X _i-X_j)m_{3/2}^2,$ where 
$X_{\Phi}$ is normalized to -1. For non-diagonal entries one has 
 $\widetilde{m}_{i\bj }^2\sim $ 
$3(X _i-X_j)m_{3/2}^2 {n_{\Phi}\cos ^2{\theta}}/
\la ^{|X _i-X_j|} .$ Similar results also follow for triscalar 
couplings.

The consequences for SFCNC are an improvement with respect to those in the
previous section. For instance, the contribution to $K\bar{K}$ mixing 
can be reduced by choosing  models\cite{DPS1} with charges $d_1=d_2$,
and the same trick is possible to avoid too much lepton flavour violation.

\section{Conclusion}
Supersymmetry is the highway connection between flavour physics at
low energies and flavour theories at the Planck scale. SFCNC 
phenomenology provide very selective constraints in this adventure.

\end{document}